\documentstyle[11pt,epsfig]{article}

\def\simg{\mathrel{\hbox{\rlap{\lower.55ex \hbox {$\sim$}}
	   \kern-.3em \raise.4ex \hbox{$>$}}}}
\def\siml{\mathrel{\hbox{\rlap{\lower.55ex \hbox {$\sim$}}
	   \kern-.3em \raise.4ex \hbox{$<$}}}}
\def\Mesz{M\'esz\'aros~}
\def\Pacz{Paczy\'nski~}
\def\beq{\begin{equation}}
\def\enq{\end{equation}}
\def\bea{\begin{eqnarray}}
\def\ena{\end{eqnarray}}

\def\bec{\begin{center}}
\def\enc{\end{center}}
\def\etal{{\it et al.}}
\def\msun{M_\odot}
\def\eps{\epsilon}

\begin{document}
\bibliographystyle{plain}

\hfill {\it Science, 291, 79 (2001)}

\bec{ \bf GAMMA-RAY BURSTS:}\enc
\bec{Accumulating Afterglow Implications, Progenitor Clues and Prospects } \enc
\smallskip\noindent
\bec {P. \Mesz$^{1,2}$ } \enc
\smallskip\noindent
{$^1$Astronomy \& Astrophysics Dpt, Pennsylvania State University,
University Park, PA 16803}\\
\smallskip\noindent
{$^2$Institute for Theoretical Physics, University of California,
Santa Barbara, CA 93106-4030}\\

{\bf
Gamma-ray bursts (GRB) are sudden, intense flashes of  gamma-rays which, 
for a few blinding seconds, light up in an otherwise fairly dark gamma-ray sky.
They are detected at the rate of about once a day, and while they are on, they 
outshine every other gamma-ray source in the sky, including the sun. 
Major advances have been made in the last three or four years, including the 
discovery of slowly fading x-ray, optical and radio afterglows of GRBs, the 
identification of host galaxies at cosmological distances, and finding evidence 
for many of them being associated with star forming regions and possibly supernovae. 
Progress has been made in understanding how the GRB and afterglow radiation arises 
in terms of a relativistic fireball shock model. 
These advances have opened new 
vistas and questions on the nature of the central engine, the identity 
of their progenitors, the effects of the environment, and their possible 
gravitational wave, cosmic ray and neutrino luminosity. The debates on 
these issues indicate that GRB remain among the most mysterious puzzles 
in astrophysics.
}

Until a few years ago, GRB were thought to be just that, bursts of gamma-rays
which were largely devoid of any observable traces at any other wavelengths.
However, a dramatic development in the last several years has been the measurement 
and localization of fading x-ray signals from some GRBs, lasting typically for days 
and making possible the optical and radio detection of afterglows, which, as fading 
beacons, mark the location of the fiery and brief GRB event. These afterglows in turn 
enabled the measurement of redshift distances, the identification of host galaxies, 
and the confirmation that GRB were, as suspected, at cosmological distances of the 
order of billions of light-years, similar to those of the most distant galaxies and 
quasars. Even at those distances they appear so bright that their energy output has 
to be of the order $10^{51}-10^{54}$ erg/s, larger than that of any other type of 
source. It is comparable to burning up the entire mass-energy of the sun in a few 
tens of seconds, or to emit over that same period of time as much energy as our entire 
Milky Way does in a hundred years.

GRBs were first reported in 1973, based on 1969-71 observations by the Vela military 
satellites monitoring for nuclear explosions in verification of the Nuclear Test Ban 
Treaty.  When these mysterious gamma-ray flashes were first detected, which did not
come from Earth's direction, the first suspicion (quickly abandoned) was that they 
might be the product of an advanced extraterrestrial civilization.  Soon, however, 
it was realized that this was a new and extremely puzzling cosmic phenomenon. 
For the next 20 years, hundreds of GRB detections were made, and frustratingly, 
they continued to vanish too soon to get an accurate angular position to permit any 
follow-up observations. The reason for this is that gamma-rays are notoriously hard 
to focus, so gamma-ray images are generally not very sharp.
\begin{figure}[ht]
\centering
\epsfxsize=2.8in
\epsfysize=2.8in
\centerline{\epsfbox{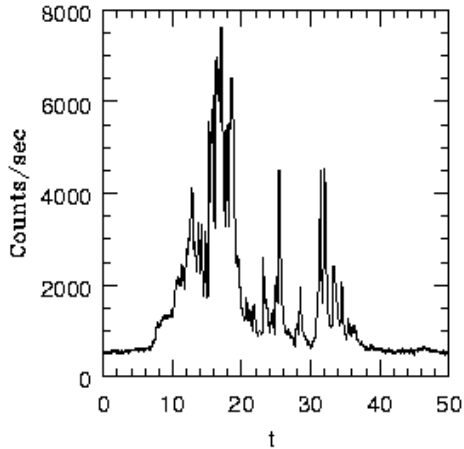} }
\caption{ 
Time Profile of a typical gamma ray burst.
The y-axis is the photon count rate in the 0.05-0.5 MeV energy,
the x-axis is the time in seconds since the burst trigger. Both before and after 
the burst trigger, no gamma-rays are detectable from the same direction \cite{fm95}.
}
   \label{fig:batselc}
\end{figure}

The next major advance occurred in 1991 with the launch of the Compton Gamma-Ray 
Observatory (CGRO), whose results have been summarized in \cite{fm95}.  The all-sky 
survey from the Burst and Transient Experiment (BATSE) onboard CGRO, which measured 
about 3000 bursts, showed that they were isotropically distributed, suggesting a 
cosmological distribution, with no dipole and quadrupole components. 
The spectra were non-thermal, the number of photons per unit photon energy varying 
typically as $N(\epsilon)\propto \epsilon^{-\alpha}$, where $\alpha \sim 1$ at low 
energies changes to $\alpha\sim 2-3$ above a photon energy $\epsilon_{0}\sim 0.1-1$ 
MeV \cite{band93}, the spectral power law dependence extending sometimes to GeV energies 
\cite{hur94}.  The durations (at MeV energies) range from $10^{-3}$ s to about $10^3$ s, 
with a roughly bimodal distribution of long bursts of $t_b \simg 2$ s and short bursts 
of $t_b \siml 2$s \cite{kou93}, and substructure sometimes down to milliseconds.
The gamma-ray light curves range from smooth, fast-rise and quasi-exponential decay,
through curves with several peaks, to variable curves with  many peaks
(Fig. \ref{fig:batselc}).
The pulse distribution is complex, and the time histories of the emission as a function
of energy can provide clues for the geometry of the emitting regions \cite{fen98}.

\begin{figure}[ht]
\centering
\vspace*{-2.0cm}
\epsfxsize=2.0in
\epsfysize=3.0in
\centerline{\epsfbox{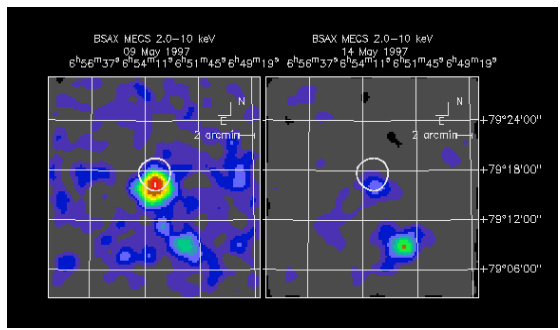} }
\caption{Beppo-SAX Narrow Field Imager pictures of the afterglow of GRB 970508 
in 2-10 keV X-rays, taken 6hours and 3 days after the burst trigger respectively,
showing the fading intensity. The white circle in the NFI image is the initial 
Wide Field Camera error box. (From L. Piro \& BeppoSAX GRB team)
}
   \label{fig:970508nfi_rot}
\end{figure}

A watershed event occurred in 1997, when the Italian-Dutch satellite Beppo-SAX 
succeeded in obtaining high resolution x-ray images\cite{cos97} of the predicted fading 
afterglow of the burst GRB 970228, followed by a number of other detections at the 
approximate rate of 10 per year (Fig. \ref{fig:970508nfi_rot}). These detections, after 
a 4-6 hour delay for processing, led to positions  accurate to about an arc-minute which 
allowed the detection and follow-up of the afterglows at optical and longer wavelengths
(e.g. \cite{jvp97}). This paved the way for the measurement of redshift distances, 
the identification of candidate host galaxies, and the confirmation that they were at 
cosmological distances \cite{metz97,kul99b}. Above 30 GRB afterglows have been located, 
with detections sometimes extending to radio \cite{fra99} and over time scales of many 
months, of which at least 25 resulted in the identification of host galaxies 
(e.g. \cite{bloom00}).

\section*{The fireball shock and afterglow scenario}

At cosmological distances the observed GRB fluxes imply energies of order of up to a solar 
rest-mass ($\siml 10^{54}$ erg), and from causality these must arise in regions whose 
size if of the order of kilometers in a time scale of the order of seconds. This implies
that an $e^\pm,\gamma$ fireball must form \cite{pac86,goo86,sp90}, which would expand 
relativistically. The difficulty with this was that a smoothly  expanding fireball would  
convert most of its energy into kinetic energy of accelerated baryons rather than into 
luminosity, and would produce a quasi-thermal spectrum, while the typical 
time scales would not explain events much longer than milliseconds.
This problem was solved with the introduction of the fireball shock model
\cite{rm92,mr93a}, based on the fact that shock waves would inevitable occur in the 
outflow, after the fireball became transparent, and these would reconvert the kinetic 
energy of expansion into nonthermal particle and radiation energy. The complicated 
light curves can be understood in terms of internal shocks \cite{rm94} in the 
outflow itself, caused by velocity variations in the outflow (c.f. also \cite{dm99}). 
This is followed by the development of a forward shock or blast wave moving into the 
external medium ahead of the ejecta, and a reverse shock moving back into the ejecta as 
the latter is decelerated by the back-reaction from the external medium 
(Fig. \ref{fig:collaps}). 

Similarly to what is observed by spacecraft in interplanetary shocks, the shocks in the 
fireball outflow are expected to be collisionless, i.e. mediated by chaotic electric 
and magnetic fields.  The minimum random Lorentz factor of protons going through the 
shocks is expected to be comparable to the bulk Lorentz factor of the flow, while that of 
the electrons may exceed this by a factor of up to the ratio of the proton to the electron 
mass. The energy of the particles can be further boosted by
diffusive shock acceleration \cite{be87} as they scatter repeatedly across the shock 
interface, acquiring a power law distribution $N(\gamma)\propto \gamma^{-p}$, where 
$p\sim 2-3$. In the presence of turbulent magnetic fields  built up behind the shocks, 
the electrons produce a synchrotron power-law radiation spectrum \cite{mr93a,rm94} similar 
to that observed \cite{band93}, while the inverse Compton scattering of these synchrotron 
photons extends the spectrum into the GeV range \cite{mrp94}. 

\begin{figure}[ht]
\centering
\epsfxsize=2.8in
\epsfysize=2.8in
\centerline{\epsfbox{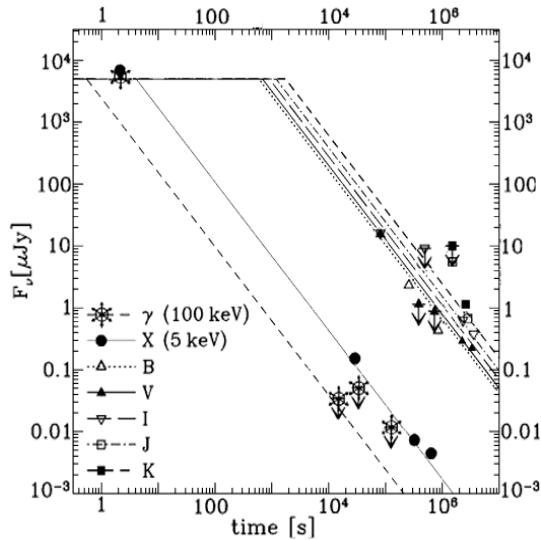} }
\caption{ 
Comparison \cite{wrm97} of the observed light curves of the afterglow of GRB 970228 
at various wavelengths with the simple blast wave model predictions \cite{mr97a}.
}
   \label{fig:sel970228}
\end{figure}

\begin{figure}[ht]
\centering
\epsfxsize=2.0in
\epsfysize=2.0in
\centerline{\epsfbox{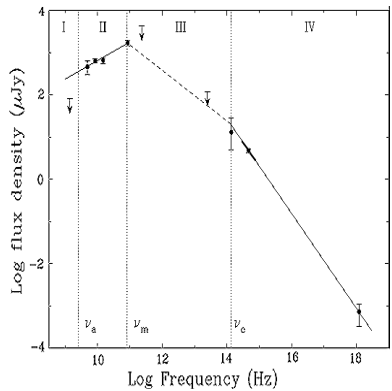} }
\caption{ 
Snapshot spectrum of GRB 970508 at $t=12$ days after the burst, compared to
a standard afterglow synchrotron shock model fit \cite{wiga99}
}
   \label{fig:spec_0508}
\end{figure}

The external shock becomes important when the inertia of the swept up external matter 
starts to produce an appreciable slowing down of the ejecta. As the fireball continues 
to plow ahead, it sweeps up an increasing amount of external matter, made up of 
interstellar gas plus possibly gas which was previously ejected by the progenitor star.
For an approximately smooth distribution of external matter, the bulk Lorentz factor of 
the fireball thereafter decreases as in inverse power of the time (which asymptotically
is $t^{-3/8}$). As a consequence, the accelerated electron minimum random Lorentz factor
and the turbulent magnetic field also decrease as inverse powers laws in time.
This implies that the spectrum softens in time, as the synchrotron peak corresponding
to the minimum Lorentz factor and field decreases \cite{rm92}, leading to the possibility
of late radio \cite{pacro93} and optical emission \cite{ka94b}.  
The GRB radiation, which started out concentrated in the $\gamma$-ray range during the 
burst, is expected to progressively evolve into an afterglow radiation which peaks in 
the X-rays, then UV, optical, IR and radio \cite{mr97a}. Detailed predictions of the
afterglow properties \cite{mr97a}, made in advance of the observations, agreed well 
with subsequent detections at these wavelengths, followed up over periods of months 
(Figs. \ref{fig:sel970228}, \ref{fig:spec_0508}). 
At a given observer frequency, after the synchrotron peak has passed through it, 
the observed photon flux also decreases as in inverse power law in time, typically
$t^{-1.2}$ or steeper. The study of GRB and afterglows 
\cite{vie97a,tav97,wrm97,rei97,spn98,wiga99,piran99} 
has provided confirmation of this generic fireball shock model of GRB, 
in agreement with the data as recently summarized in a review \cite{jvp00}.
An important check on the model came from the detection of diffractive scintillation
in the radio afterglow of GRB970508, which provided a direct determination of the
source size and a direct confirmation of relativistic source expansion 
\cite{goo97,wfk98}.

One issue raised by the large redshifts \cite{kul99b}, is that the measured 
$\gamma$-ray fluences imply a total photon energy of order $10^{52}-10^{54}
(\Omega_\gamma /4\pi)$ ergs, where $\Omega_\gamma$ is the solid angle into 
which the gamma-rays are beamed. For a solar mass object, this implies that an
unusually large fraction of the energy is converted into $\gamma$-ray photon energy.
A beamed jet would alleviate the energy requirements, and some observational 
evidence suggests the presence of a jet \cite{kul99a,fru99,cas99,sari+99}.
Whether a jet is present or not, such energies are in principle achievable for
bursts arising from stellar progenitors, but a poorly understood issue is how this 
energy is converted into an ultrarelativistic, and possibly collimated bulk outflow.

An observation which attracted much attention was the discovery \cite{ak99} of a prompt
and extremely bright ($m_v\sim 9$) optical flash in the burst GRB 990123, 15 seconds after 
the GRB started (and while it was still going on). This is generally interpreted 
\cite{sp99,mr97a} as the radiation from the reverse component of the external shock. Such 
bright prompt flashes, however, may be rare, since they have not so far been detected from 
other bursts.  
Two other noteworthy developments are the possibility of a relationship between the 
differential time lags for the arrival of burst pulses at different energies and the 
luminosity \cite{norris00}, and between the degree of variability or spikyness of the 
gamma-ray light curve variability and the luminosity \cite{fen00,rei00}, based on data 
for bursts where an optical redshift allows a determination of the luminosity, under 
the assumption of isotropy.  These correlations are tentative so far, but if confirmed 
they could be used for independently estimating the redshift of a GRB.

\section*{Progenitors and Environment}

\begin{figure}[ht]
\centering
\epsfig{file=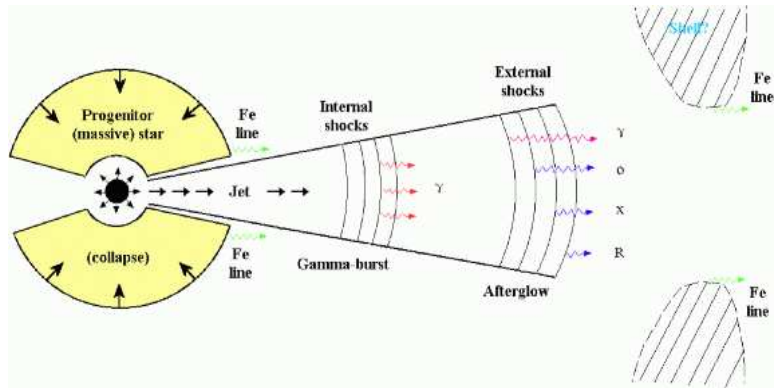,width=4.0in,height=2.0in} 
\caption{
Schematic GRB from a massive stellar progenitor, resulting in a relativistic jet which 
undergoes internal shocks producing a burst of $\gamma$-rays and (as it decelerates 
through interaction with the external medium) an external shock afterglow which leads 
successively to $\gamma$-rays, X-rays, optical and radio. Iron lines may arise
from X-ray illumination of a pre-ejected shell (e.g. supernova remnant)\cite{piro00}
or from continued X-ray irradiation of the outer stellar envelope \cite{rm00}.
}
   \label{fig:collaps}
\end{figure}

The progenitors of GRB are not well identified so far. The current view of a majority of 
researchers is that GRBs arise in a very small fraction ($\sim 10^{-6}$) of stars 
which undergo a catastrophic energy release event toward the end of their evolution. 
One class of candidates involves massive stars whose core collapses 
\cite{woo93,pac98,fwh00}, probably in the course of merging with a companion,
often referred to as hypernovae or collapsars \cite{note:progdef}. 
Another class of candidates consists of neutron star (NS) binaries or neutron 
star-black hole (BH) binaries, \cite{pac86,goo86,eic89,mr97b} which lose orbital angular 
momentum by gravitational wave radiation and undergo a merger. 
Both of these progenitor types are expected to have as an end result the formation 
of a few solar mass black hole, surrounded by a temporary debris torus whose accretion can 
provide a sudden release of gravitational energy, with similar total energies \cite{mrw98}, 
sufficient to power a burst. An $e^\pm ,\gamma$ fireball arises from the enormous 
compressional heating and dissipation associated with the accretion, possibly
involving a small fraction of baryons and magnetic fields in excess of $10^{15}$ Gauss, 
which can provide the driving stresses leading to the relativistic expansion. This 
fireball may be substantially collimated, if the progenitor is a massive star, where
an extended, fast-rotating envelope can provide a natural escape route or funnel for the 
fireball along the rotation axis (Fig. \ref{fig:collaps}).
Other possible alternatives include the formation from a stellar collapse of a fast-rotating 
ultra-high magnetic field neutron star \cite{us94,tho94,hs99}, or the tidal disruption of 
compact stars by $10^5-10^6 \msun$ black holes \cite{bo00}.

Observation related to the possible progenitors are restricted, so far, to the class 
of long bursts (of $\gamma$-ray durations $t_b \sim 10-10^3$ s), because BeppoSAX 
is mainly sensitive to bursts longer than about 5-10 s. For these long bursts, the fading 
x-ray and optical afterglow emission is predominantly localized within the optical image 
of the host galaxy.  In most cases it is offset from the center, but in a few cases (out of 
a total of about twenty) it is near the center of the galaxy \cite{bloom00}. This is in 
disagreement with current simple calculations of NS-NS mergers which suggest that high
spatial velocities would take these binaries, in more than half of the cases, outside of 
the confines of the host galaxy before they merge and produce a burst. These calculations,
however, are uncertain, since they are sensitive to a number of poorly known parameters 
(e.g distribution of initial separations, etc).
On the other hand, theoretical estimates \cite{fwh00} suggest that NS-NS and NS-BH mergers 
will lead to shorter bursts ($\siml 5 s$), beyond the capabilities of Beppo-SAX but expected 
to be detectable with the recently launched HETE-2 spacecraft \cite{hete00} and the Swift
multi-wavelength GRB afterglow mission \cite{swift00} now under construction.

For the long burst afterglows localized so far, the host galaxies show signs of ongoing 
star formation activity, necessary for the presence of young, massive progenitor stars. 
Such stars generally form in dense gaseous clouds, for which there is some independent
evidence from the observation of 0.5-2 keV absorption in the x-ray afterglow spectra, 
attributed to metals in a high column density of gas in front of the burst \cite{gawi00}. 
X-ray atomic edges and resonance absorption lines are expected to be detectable from the 
gas in the immediate environment of the GRB, and in particular from the remnants of a
massive progenitor stellar system \cite{mr98b,weth00,botfry00}. Observations with the 
Chandra ACIS x-ray spectrographic camera and with BeppoSAX have provided evidence, with
a moderate confidence level, for iron K-$\alpha$ line and edge features in at least two 
bursts \cite{piro00,amati00}.  The observed frequency of the iron lines appear displaced 
from the laboratory frequency, as expected from the Doppler shift caused by the expansion 
of the universe, in agreement with the redshift measured in optical lines from 
the host galaxy. 

One possible interpretation of the iron lines is that x-rays from the afterglow illuminate
an iron-enriched supernova remnant situated outside the burst region, leading to iron 
recombination line emission (Fig. \ref{fig:collaps}).  This would require the supernova 
explosion to have occurred days or weeks before the burst, associated with the same 
progenitor\cite{piro00,vie00,note:Fe}. There is independent support that, at least in 
some bursts, a supernova may be involved \cite{gal98-SN,jvp99,whee99}. This may have 
contributed to an otherwise unexplained bump and reddening in the optical light curve 
after several weeks, and similar reddened bumps have been reported in at least two other 
bursts. The presence of iron line features would strongly suggest a massive stellar 
progenitor \cite{piro00}, but the details remain model dependent. Even without a 
pre-ejected supernova shell, a continued decaying X-ray emission from the GRB outflow 
impacting the outer stellar envelope \cite{rm00,note:Fe} may explain the iron lines. 

The simple picture of an origin in star-forming regions, at least for the long
($t_b \simg 5$ s) bursts, is complicated by the fact that the observed optical 
absorption is less than expected for the corresponding x-ray absorption.
Also, standard afterglow model fits indicate an ambient gas density generally lower than 
that expected in star-forming clouds \cite{gawi00}. These contradictions, however,
may be reconciliable, e.g. through dust sublimation by x-ray/UV radiation, or the blowing 
out of a cavity by a progenitor wind.

While it is unclear whether there is one or more classes of GRB progenitors,
e.g. corresponding to short and long bursts, there is a general consensus that they 
would all lead to the generic fireball shock scenario. Much of the current effort is 
dedicated to understanding the different progenitor scenarios, and trying to determine how 
the progenitor and the burst environment can affect the observable burst and afterglow 
characteristics.

\section*{Galactic Hosts and Cosmological Setting}

\begin{figure}[ht]
\centering
\epsfig{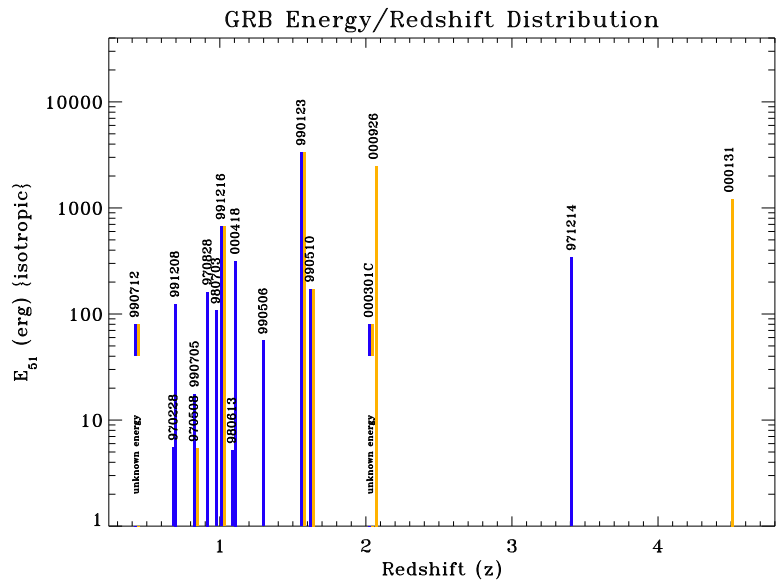}
\caption{
The observed energy-redshift relation for 17 GRBs with optical
spectroscopic redshifts as of 10/27/2000.  Blue (orange) denotes that the redshift 
was found with emission (absorption) lines from the presumed host galaxy.  
The energy is derived from the gamma-ray fluences reported and assumes
that the GRB emitted energy isotropically. (Courtesy of J.S. Bloom and 
the Caltech GRB group).
}
   \label{fig:figz}
\end{figure}

For the long GRB afterglows localized so far, a host galaxy has been found in most cases 
(a growing number, over 20 out of 30 optically identified). The GRB host galaxies 
are typically of low mass, and have the blue color and atomic spectral 
lines indicative of active star formation \cite{bloom00}. The redshifts of the hosts, 
with one exception, are in the range $0.43 \siml z \siml 4.5$ (Fig. \ref{fig:figz}), i.e., 
comparable to that of the most distant objects detected in the Universe (about $10^{10}$ 
light-years). The observed number of bursts per unit photon flux can be fitted by 
cosmological distribution models, with a somewhat better fit if one assumes that the burst 
rate scales proportionally to the observed star-formation rate as a function of redshift 
\cite{wi98,tot99,bn00}. The spread in the inferred luminosities (Fig. \ref{fig:figz}) is 
too broad to allow the use of GRB as standard candles for the purposes of testing 
cosmological models \cite{maomo98}. This spread in the inferred luminosities obtained 
under the assumption of isotropic emission may be reduced if most GRB outflows are 
jet-like, because in this case the measured flux is more intense when observed closer 
to the jet axis, due to an increased Doppler boost.

The bursts for which redshifts are known are bright enough to be detectable, in principle,
out to much larger distances than those of the most luminous quasars or galaxies detected 
at present \cite{lr00}. Within the first minutes to hours after the burst, the optical 
light from afterglows is known to have a range of visual magnitudes $m_v \sim 10-15$, far 
brighter than quasars, albeit for a short time. Thus, promptly localized GRB could serve 
as beacons which, shining through the pregalactic gas, provide information about much 
earlier epochs in the history of the Universe. The presence of iron or other x-ray 
lines provides an additional tool for measuring GRB distances, which may be valuable for 
investigating the small but puzzling fraction of bursts which have been detected only in 
X-rays but not optically, perhaps due to a high dust content in the host galaxy.

The newly launched HETE spacecraft \cite{hete00} is expected to yield localizations for 
about 30 bursts per year, and up to 200-300 per year are expected to be localized 
with the Swift spacecraft \cite{swift00} due for launch in 2003.  Swift will be equipped 
with $\gamma$-ray, x-ray and optical detectors for on-board follow-up, and capable of
relaying to the ground arc-second quality burst coordinates within less than a minute 
from the burst trigger, allowing even mid-size ground-based telescopes to obtain prompt
spectra and redshifts. This will permit much more detailed studies of the burst environment,
the host galaxy, and the intergalactic medium between galaxies. 
The diffuse gas around a GRB is expected to produce time-variable optical/UV
atomic absorption lines in the first minutes to hours after a burst \cite{pl98}, and
additional hydrogen Lyman $\alpha$ absorption from intervening newly formed galaxies would 
be detectable as the GRB optical/UV continuum light shines through them \cite{lb00}.
While the starlight currently detected is thought to come mostly from later, already 
metal-enriched generations of star formation, GRB arising from the earliest generation of 
stars may be detectable; and if this occurs before galaxies have  gravitationally assembled, 
it would provide a glimpse into the pregalactic phase of the Universe.

\section*{Cosmic rays, neutrinos and gravitational waves } 

There are other, as yet unconfirmed, but potentially interesting manifestations of GRBs.
The same shocks which are thought to accelerate the electrons responsible for the 
non-thermal $\gamma$-rays in GRB should also accelerate protons.  Both the  internal 
and the external reverse shocks are mildly relativistic, and are expected to lead to 
relativistic proton energy spectra of the form $d N_p/d \eps_p \propto \gamma_p^{-2}$. 
The maximum proton energies achievable in GRB shocks are $E_p\sim 10^{20}$ eV, comparable 
to the highest energies measured with large cosmic ray ground arrays, e.g. \cite{ag99}. 
For this, the acceleration time must be shorter than both the radiation or adiabatic loss 
time and the escape time from the acceleration region. The resulting constraints on the 
magnetic field and the bulk Lorentz factor \cite{waxman95} are close to those required to 
obtain efficient gamma-ray emission at $\sim 1$ MeV. If the accelerated electrons which 
produce the $\gamma$-rays and the protons carry a similar fraction of the total energy, 
the GRB cosmic ray energy production rate at $10^{20}$ eV throughout the universe is of 
order $10^{44}$ erg/Mpc$^3$/yr, comparable to the observationally required rate from 
$\gamma$-ray observations and from the observed diffuse cosmic ray flux
\cite{waxman95,der00} (c.f. \cite{ste00}). These numbers depend to some extent on 
uncertainties in the burst total energy and beaming fraction, as well as on the 
poorly constrained burst rate evolution with redshift.

The accelerated protons can interact with the fireball photons, leading to 
charged pions, muons and neutrinos. This reaction peaks at the energy threshold for 
the photo-meson $\Delta$ resonance. For internal shocks producing observed 1 MeV
photons this implies $\simg 10^{16}$ eV protons, and neutrinos with $\sim 5\%$ 
of that energy, $\eps_\nu\simg 10^{14}$ eV.  Above this threshold, the fraction of the 
proton energy lost to pions is $\sim 20\%$ for typical fireball parameters, and the
typical spectrum of neutrino energy per decade is flat, $\eps_\nu^2 \Phi_\nu \sim$ constant 
\cite{wb97}.  Synchrotron and adiabatic losses limit the muon lifetimes \cite{rm98}, 
leading to a suppression of the neutrino flux above $\eps_\nu \sim 10^{16}$ eV. 
Another copious source of target photons in the UV is the afterglow reverse shock, 
for which the resonance condition requires higher energy protons leading to neutrinos 
of $10^{17}-10^{19}$ eV \cite{wb99}.  
These neutrino fluxes are expected to be detectable above the atmospheric neutrino 
background with the planned cubic kilometer ICECUBE Cherenkov detector \cite{halzen00}.

Another mechanism for neutrino production in GRB is inelastic nuclear collisions.
Whereas photo-pion interactions lead to higher energy neutrinos and provide a direct 
probe of the shock proton acceleration as well as of the photon density, inelastic  
proton-neutron collisions  may occur even in the absence of shocks, leading to charged 
pions and neutrinos \cite{der99} with lower energies than those from photo-pion 
interactions. Provided the fireball has a substantial neutron/proton ratio, as expected in 
most GRB progenitors, the inelastic process is most intense when the nuclear scattering 
time scale becomes comparable to the expansion time scale, at which point the relative 
velocities of the nuclei become large enough to collide inelastically, resulting in charged
pions and neutrinos \cite{bm00}. Inelastic collisions can also occur in fireball outflows
with transverse inhomogeneities in the bulk Lorentz factor \cite{mr00}. The typical 
neutrino energies are in the 1-10 GeV range, which could be detectable for a sufficiently 
close  photo-tube spacing in Km$^3$ detectors, in coincidence with observed GRBs.

The photo-pion and inelastic collisions responsible for the ultra-high energy neutrinos 
will also lead to neutral pions and electron-positron pair cascades, resulting in GeV to 
TeV energy photons. A tentative $\simg 0.1$ TeV detection of a GRB has been reported with 
the water Cherenkov detector Milagrito \cite{atk00}. Other large atmospheric Cherenkov 
detectors, as well as planned space-based large area solid state detectors such as GLAST 
\cite{glast00} will be able to measure photons in this energy range, which would be 
coincident with the neutrino pulses and the usual MeV $\gamma$-ray event. Their 
detection would provide important constraints on the emission mechanism of GRBs.

GRB are also expected to be sources of gravitational waves. A time-integrated luminosity
of the order of a solar rest mass ($\sim 10^{54}$ erg) is predicted from merging NS-NS and 
NS-BH models, while the luminosity from collapsar models is less certain, but estimated 
to be lower. Calculations \cite{finn00} of the rates of gravitational wave events 
detectable by the Laser Interferometric Gravitational Wave Observatory (LIGO, currently 
under construction) from compact binary mergers, in coincidence with GRBs, has been 
estimated at a few/year for the initial LIGO, and up to 10-15/year after the upgrades 
planned 2-4 years after first operations.  The observation of such gravitational waves 
would be facilitated if the mergers involve observed GRB sources; and conversely, 
it may be possible to strengthen the case for (or against) NS-NS or NS-BH progenitors of 
GRB if gravitational waves were detected (or not) in coincidence with some bursts.

In conclusion, our understanding of GRB has come a long way since their discovery 
almost 30 years ago, but these enigmatic sources continue to offer major puzzles and 
challenges.  Several new space missions and ground experiments dedicated to GRB studies 
will come on-line in the near future, which should answer many of the questions
discussed here. If past experience is any guide, they will also undoubtedly come up 
with new surprises and challenges to look forward to. \cite{note:ack}

\bigskip
\def\bitm{\bibitem}



\end{document}